\documentclass{osa-article}

\journal{oe}
\usepackage[utf8]{inputenc}
\articletype{Research Article}
\begin{document}

\title{Reflectionless Plasmonic Right-Angled Waveguide Bend and Divider Using Graphene and Transformation Optics}

\author{Samaneh Pakniyat,\authormark{1,2,*} Shahrokh Jam,\authormark{1} Alireza Yahaghi\authormark{3} and George W Hanson \authormark{2}}
\address{\authormark{1}Shiraz university of technology, Department of Electrical Engineering, Shiraz, Iran\\ 
\authormark{2} University of Wisconsin-Milwaukee, Department of Electrical Engineering, Milwaukee, Wisconsin, USA\\
\authormark{3} Shiraz university, Department of Electrical Engineering, Shiraz, Iran}
\email{\authormark{*}s.pakniyat@sutech.ac.ir} %% email address is required
%\email{\authormark{*}pakniyat@uwm.edu} %% email address is required

%%%%%%%%%%%%%%%%%%% abstract %%%%%%%%%%%%%%%%

\begin{abstract}
In this work, a plasmonic right-angled waveguide bend and divider are proposed. Using the Transformation Optics (TO) approach the transformation media of a bend and a T-shaped divider are obtained. Such media with continuous refractive index are realized with the help of graphene in the terahertz frequency range, key to effectively guiding the surface plasmon polariton (SPP) propagation on the 90 degree curves. Components with such capability are promising for THz device applications.
\end{abstract}

%%%%%%%%%%%%%%%%%%%%%%%%%%  body  %%%%%%%%%%%%%%%%%%%%%%%%%%
\section{Introduction}
\setcounter{page}{1} Transformation Optics (TO) is a mathematical method
based on the invariance of the Maxwell's equations under coordinate
transformation \cite{pendry,Leo06}. According to Fermat's principle, a wave
in a medium with non-uniform refractive index deviates from a straight line
and propagates along a curved path, the shortest optical path \cite{Fermat}.
With the help of the Transformation Optics approach, the parameters of such
a medium can be analytically obtained to effectively direct the
electromagnetic wave. In TO approach, the idea is to map a virtual domain
with desirable propagation behavior into a physical domain. Then, the
coordinate transformation is interpreted as the parameters of the material
filling the physical domain \cite{Leo09}. As a result, electromagnetic wave
propagation in the physical domain has the desired characteristics.

The TO method has been broadly used in the design of novel electromagnetic
devices, namely invisibility cloaks, polarization rotators, polarization
splitters, beam concentrators, and beam convertors \cite{Cai, Huang, Rahm,
Wang, Jiang08, Jiang09, Kwon, Chen}. It is also usable in the optimization
of classical electromagnetic devices such as antennas, waveguides, and
lenses \cite{Tichit, Tang, Aghanejad}. However, the practical realization of
transformation optics media governed by TO is still lagging. The main subject of our work is implementation of transformation media of the
right-angle bend and T-shape divider at infrared frequencies, using a finite-width plasmonic graphene waveguide with variable bias. In the millimeter and microwave frequency range, different techniques such as graded index (GRIN) photonic crystal \cite{Mola, gilarlue2019photonic}, subwavelength grating metamaterial \cite{badri2020silicon}, and substrate thickness variation \cite{badri2020ultra} have been used to implement a TO gradient index. In this work, we take advantage of graphene to design a reflectionless plasmonic bend and divider. 

Graphene can serve as a suitable palsmonic platform for realization of transformation media in the infrared frequency range, thanks to its unique optoelectronic properties. It is a two-dimensional dispersive material which supports SPPs. Due to low loss, high confinement
of SPPs, and more importantly tunable conductivity features, graphene is a
more feasible platform than the noble metals (e.g. gold, silver) in the low
THz and infrared frequency ranges \cite{Vaki2}. The graphene conductivity is
a function of different parameters such as the chemical potential,
temperature and frequency. Since it is a two-dimensional material, the substrate features
affect carrier density on the graphene surface, which also affect the
chemical potential and the graphene conductivity \cite{Vaki2}. This provides
more degrees of freedom in design and optimization of TO graphene-based
waveguides.

A few TO media have been previously proposed to be implemented by graphene. In \cite{Morgan}, the potential of applying graphene has been only examined theoretically for a TO collimator and a coupler. In \cite{Huidobro, Xia}, using the TO approach a graphene sheet with uneven substrate has been proposed as an efficient grating to enhance the light energy coupling to the surface plasmons.  A graphene plasmonic metasurfaces and an anisotropic graphene metamaterial have been designed for a TO fish-eye lens and a bend, respectively in Refs.\cite{Zeng} and \cite{chen2016tunable}. In \cite{Vakil1, Vaki2}, a TO Luneburg lens has been  designed by creating several concentric rings with different Fermi levels. It is concerned with manipulating the propagation of 2D graphene SPPs (2DGSPPs, modes of infinite graphene sheet) to alter the cylindrical wavefrount to the plane wavefrount.

The state of the art is to control the propagation of the waveguide (ribbon) graphene SPPs (GSPPs) on the finite-width graphene strips, which are more practical. 
To the best of our knowledge, our work is the first study on controlling the waveguide GSPPs on nanoribbons with a sharp bend using the TO approach. We propose a low profile discrete model to realize the TO transformation media of a 90 degree waveguide bend using graphene. Due to utilizing graphene, good SPP propagation properties can be obtained. This design also provides a wide bandwidth, unlike graphene metamaterial designs which are resonance-based. Moreover, such graphene components are usable in graphene-based integrated circuits \cite{romagnoli2018graphene}. They can be utilized to shrink the size of infrared and optical circuits and to improve their performance.

It should be noted that in addition to using TO, other techniques such as applying nano-ribbon subwavelength width
 \cite{zhu2013bends}, and excitation of nonreciprocal edge GSPPs by applying static magnetic bias \cite{CrI3} can be applied to design reflection-less graphene bend structures. However, they work only for edge GSPP, not waveguide GSPP. Furthermore, in \cite{lu2013flexible}, based on a critical bend radius computation, the 2DGSPPs propagate with no reflection on a  folded graphene sheet, which is a bulky configuration. In \cite{Vaki2,lu2013flexible}, a Y-shape graphene splitter has been reported, in which a forbidden and allowed propagation path is defined by adjusting two different Fermi level. This kind of design works only for small junction angles.

In this work, we apply the inverse conformal transformation, a subclass of the TO method, which has been presented in \cite{Turpin, Ma} for bend and divider to obtain the transformation media of the nanostructures. The TO application is not limited to a particular frequency range; it can be applied in design of infrared and optical devices as well as microwave devices. The conformal mapping provides isotropic parameters, leading to less complex fabrication.

In the following, first using TO and the inverse conformal mapping, we
determine the transformation media of a nano right-angled bend and a
T-shaped divider. Then, we realize such ideal transformation media by
applying a discrete model based on graphene. The chemical potential is
chosen as a variable factor to actualize the non-uniform refractive index
profile. Our main interest is to guide the first waveguide surface mode on
the bend structures with finite width. Therefore,  the effect of the
chemical potential on the propagation properties of this mode in a graphene
ribbon is investigated to derive the refractive index of this particular
mode. Finally, two THz reflection-less waveguide bends (a right-angled bend
and a T-shaped divider) are designed, wherein the surface plasmon polaritons
are effectively guided and do not have any reflection upon encountering the
waveguide curves. The optimal performances of the proposal designs are
demonstrated using a full-wave simulation.

\section{DESIGN METHOD}
\subsection{Transformation media of a bend waveguide and a T-junction}

\begin{figure}
\centering\includegraphics[width=10cm]{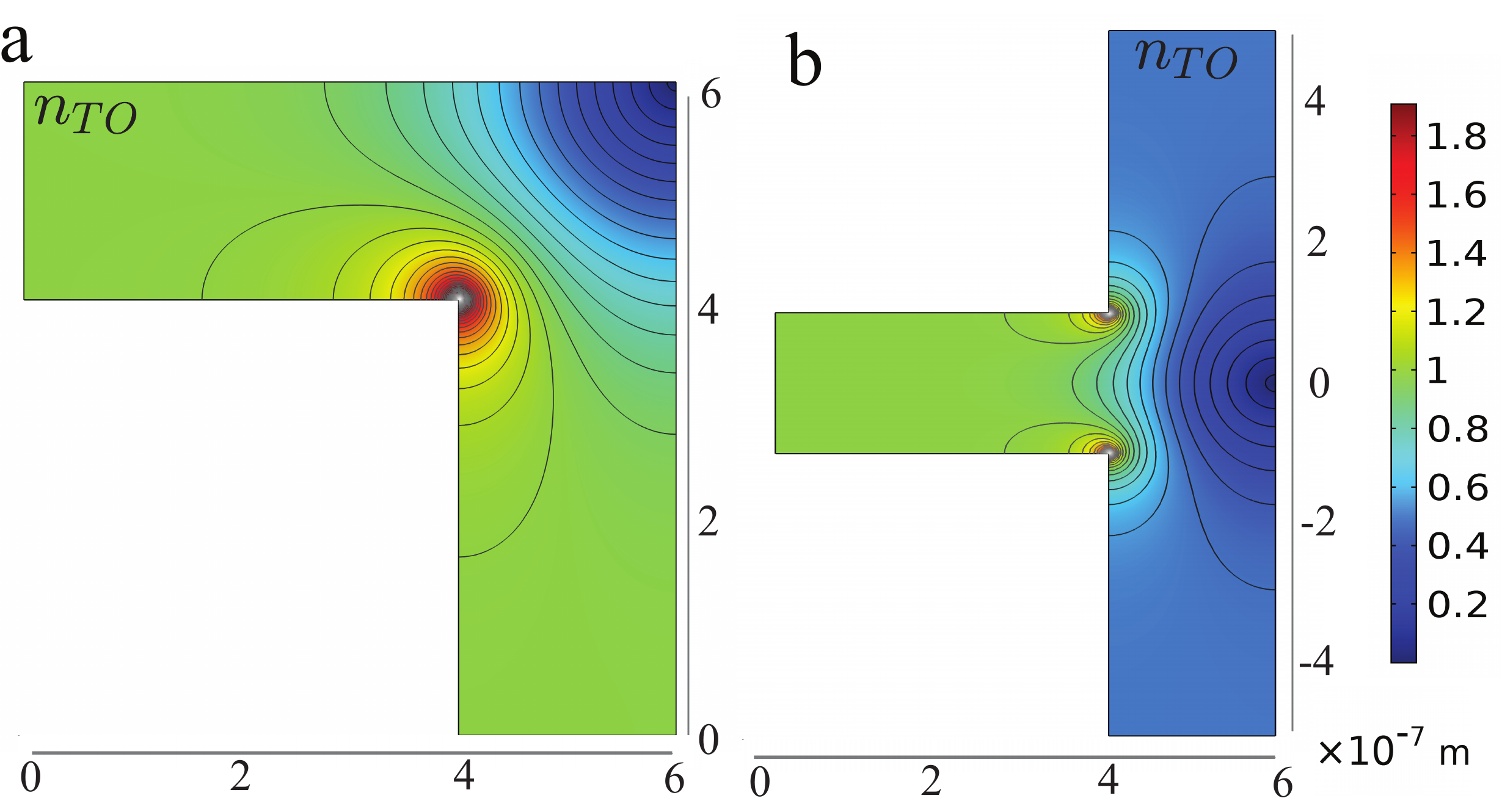}
  \caption{The continuous refractive index profile of (a) a TO bend and (b) a TO divider. The patterns sketched by the black lines indicate the regions with almost uniform refractive index.  
  }
 \label{nTO}
\end{figure}

According to Transformation Optics, under a transmission from a virtual domain ($x%
{\acute{}}%
,y%
{\acute{}}%
,z%
{\acute{}}%
$) to a physical domain ($x,y,z$), the permittivity and permeability of the
material in the physical domain are calculated by $\varepsilon _{p}=\Lambda
\varepsilon _{r}\Lambda ^{T}/\left\vert \Lambda \right\vert $ and $\mu
_{p}=\Lambda \mu _{r}\Lambda ^{T}/\left\vert \Lambda \right\vert $, where $\varepsilon _{r}$ and $\mu _{r}$ are the material
parameters in the virtual domain, and $\Lambda $ is the Jacobian
transformation matrix as $\Lambda \equiv \partial (x%
%TCIMACRO{\U{b4}}%
%BeginExpansion
{\acute{}}%
%EndExpansion
,y%
%TCIMACRO{\U{b4}}%
%BeginExpansion
{\acute{}}%
%EndExpansion
,z%
%TCIMACRO{\U{b4}}%
%BeginExpansion
{\acute{}}%
%EndExpansion
)/\partial (x,y,z)$ \cite{Leo09}. We apply an inverse conformal mapping with the coordinate
transformations in the form of $x=u(x%
%TCIMACRO{\U{b4}}%
%BeginExpansion
{\acute{}}%
%EndExpansion
,y%
%TCIMACRO{\U{b4}}%
%BeginExpansion
{\acute{}}%
%EndExpansion
)$, $y=v(x%
%TCIMACRO{\U{b4}}%
%BeginExpansion
{\acute{}}%
%EndExpansion
,y%
%TCIMACRO{\U{b4}}%
%BeginExpansion
{\acute{}}%
%EndExpansion
)$ and $z=z%
%TCIMACRO{\U{b4}}%
%BeginExpansion
{\acute{}}%
%EndExpansion
$. 
\begin{figure}
\centering\includegraphics[width=10cm]{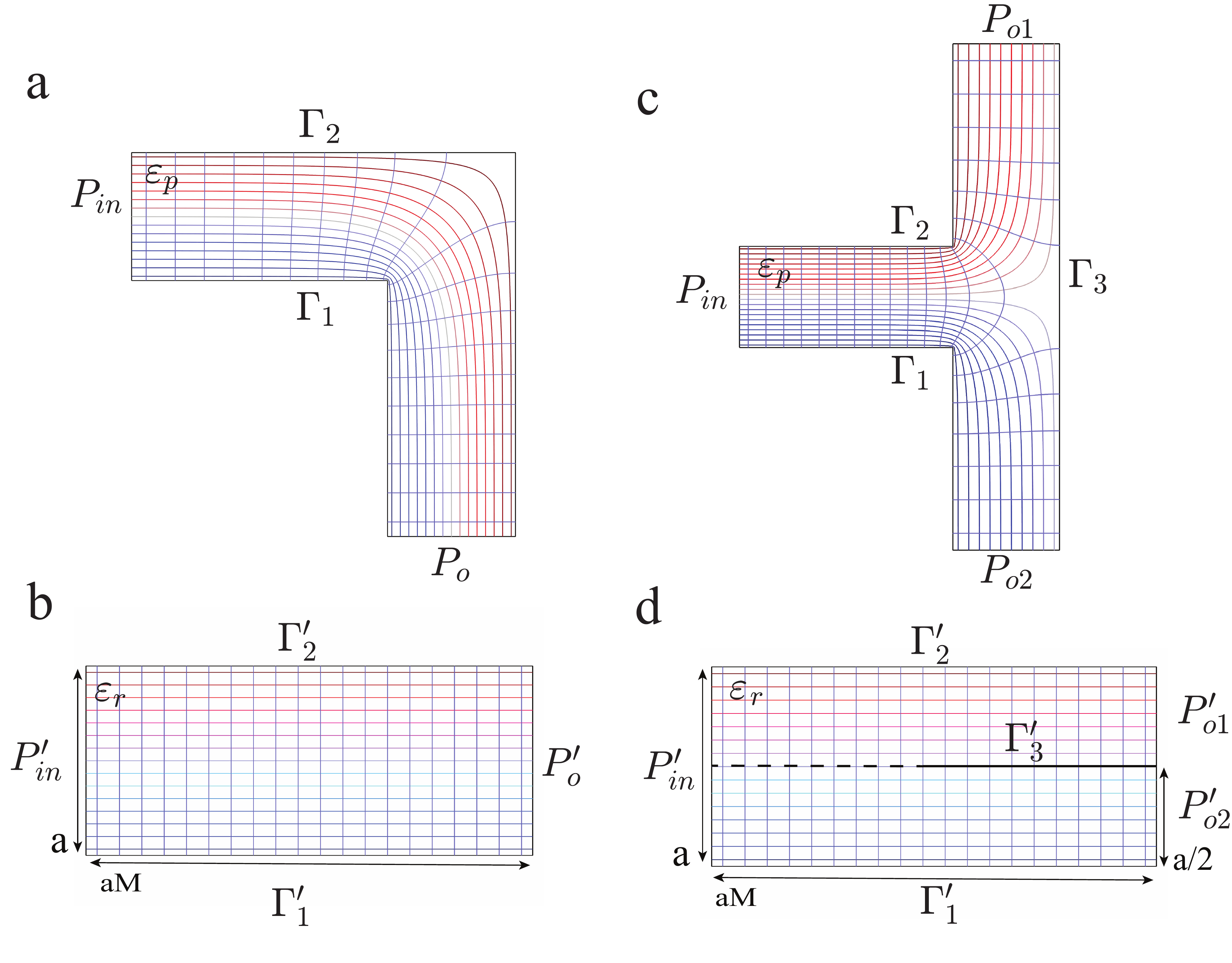}
  \caption{The physical and virtual domain of (a,b) a right-angled waveguide bend (c,d) a T-shaped divider. In the conformal transformation, the boundaries with similar notation (with and without prime superscript) are mapped to each other. The transformation and Cartesian coordinates are illustrated by colored lines inside the geometries. 
  }
 \label{PV}
\end{figure}
This corresponds to solving two Laplace's equations of \ $u_{xx}+$\ $u_{yy}=0$
and $v_{xx}+$\ $u_{yy}=0$ with Dirichlet and Neumann boundary conditions$.$
As suggested in Ref. \cite{Ma}, a right-angled waveguide bend and a straight waveguide shown in Fig. \ref{PV}a and \ref{PV}b are considered as the physical and virtual domains, respectively. Following\cite{Ma}, the Dirichlet and Neumann boundary conditions of this TO problem are 
\begin{eqnarray}
\left. u\right\vert _{\Gamma _{1}} &=&0,\left. u\right\vert _{\Gamma
_{2}}=a,\left. \hat{n}\cdot \bigtriangledown u\right\vert _{all-ports}=0 \nonumber \\
\left. v\right\vert _{P_{in}} &=&0,\left. v\right\vert _{P_{o}}=aM,\left. 
\hat{n}\cdot \bigtriangledown v\right\vert _{\Gamma _{1},\Gamma _{2}}=0,
\end{eqnarray}%
where $a$ is a scaling factor, $M$ is the conformal module given by $%
M=1/a\int_{\Gamma _{2}}\left\vert \partial u/\partial n\right\vert ds$ \cite{Henrici},
and $\hat{n}$ is a normal vector to the boundaries.
For the second design, a T-shaped divider and a three port straight waveguide
shown in Fig. \ref{PV}c and \ref{PV}d are chosen as the physical and virtual domains, respectively, with Dirichlet and Neumann boundary conditions as \cite{Ma}

\begin{eqnarray}
\left. u\right\vert _{\Gamma _{1}} &=&0,\left. u\right\vert _{\Gamma
_{2}}=a,\left. u\right\vert _{\Gamma _{3}}=a/2,\left. \hat{n}\cdot
\bigtriangledown u\right\vert _{all-ports}=0 \nonumber \\
\left. v\right\vert _{P_{in}} &=&0,\left. v\right\vert _{P_{o1}, P_{o2}}=aM,\left. \hat{n}\cdot \bigtriangledown v\right\vert
_{all-walls}=0.
\end{eqnarray}
The Laplace's
equations can be solved by a normal PDE solver using COMSOL Multiphysics software.
In the inverse conformal transformation the physical domain permittivity is modified as $\varepsilon _{p}=
\varepsilon _{r}/\left\vert \Lambda 
%TCIMACRO{\U{b4}}%
%BeginExpansion
{\acute{}}%
%EndExpansion
^{-1} \right\vert $  where $\Lambda 
%TCIMACRO{\U{b4}}%
%BeginExpansion
{\acute{}}%
%EndExpansion
\equiv \partial (x,y,z)/\partial (x%
%TCIMACRO{\U{b4}}%
%BeginExpansion
{\acute{}}%
%EndExpansion
,y%
%TCIMACRO{\U{b4}}%
%BeginExpansion
{\acute{}}%
%EndExpansion
,z%
%TCIMACRO{\U{b4}}%
%BeginExpansion
{\acute{}}%
%EndExpansion
)$ is the Jacobian matrix of the inverse transmission and $\Lambda =(\Lambda 
%TCIMACRO{\U{b4}}%
%BeginExpansion
{\acute{}}%
%EndExpansion
)^{-1}$ \cite{Ma}. The refractive indices of two TO transformation media are calculated by 
\begin{eqnarray}
n_{TO}=\sqrt{\varepsilon _{p}}=\sqrt{\varepsilon _{r}(u_{x}^{2}+v_{x}^{2})}
\end{eqnarray}%
assuming $\mu _{p}=1$. Figure \ref{nTO}a and \ref{nTO}b show the refractive index profiles of a TO bend and a TO divider.
 As shown, there is a wide range of index variations. The inhomogeneity of the refractive index results in the locally variable phase velocity, leading to
the guidance of the electromagnetic wave in the bends and reduction of the
reflection.
To aid in implementing the obtained index profiles, the refractive index profiles with continuous variation are divided into several regions in which the refractive index is almost uniform. The regions form different patterns as shown in Fig. \ref{nTO}a,b with black lines.

\subsection{ SP modes of a graphene ribbon}

\begin{figure}
\centering\includegraphics[width=10cm]{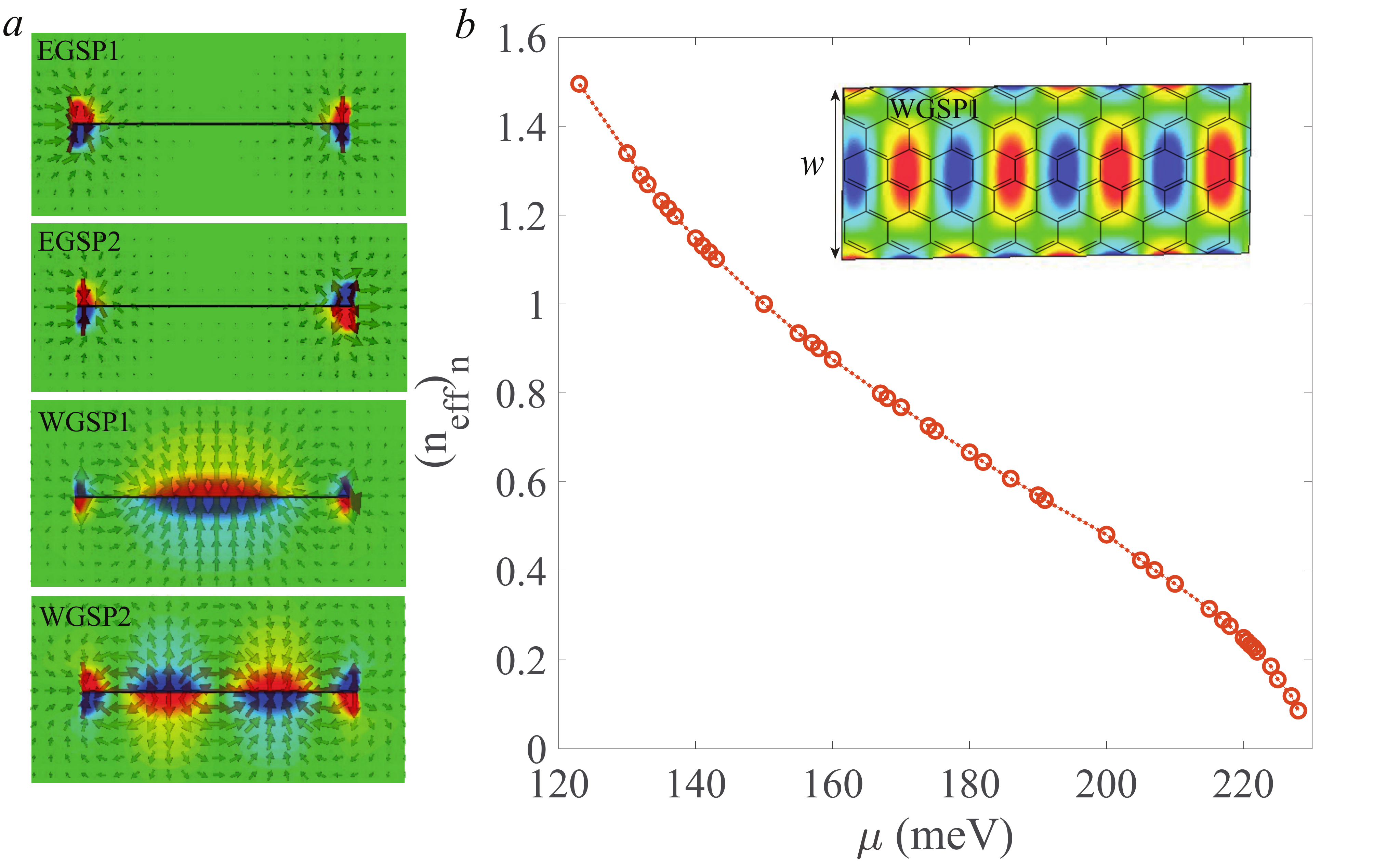}
  \caption{(a) The edge and waveguide modes in a graphene ribbon with width $w=200$nm. (b) Normalized effective refractive index, ($n_{eff})_n$, of the first waveguide graphene surface mode (WGSPP1) versus the chemical potential. $n_{\text{eff}=\beta/k_0}$.
  }
 \label{wgsp}
\end{figure}

Graphene is a two-dimensional dispersive material which supports SPPs. The dispersion
properties of TE and TM SPP modes propagating in a free-standing graphene sheet as well as
a graphene ribbon have been widely studied \cite{Hanson1, Hanson2, Mikhailov, Nikitin}. Here, we need to extract the effective refractive index of TM SPP mode at various the chemical potential for operating frequency. Considering a graphene ribbon with the width $w,$ Fig. \ref{wgsp}a shows the first four propagating modes of the ribbon, obtained by a modal analysis using
CST\ software. In the simulation, graphene is modeled by a
single suspended layer with the effective permittivity $\varepsilon
_{g}\cong (-\sigma _{g,i}+i\sigma _{g,r})/\omega \Delta$, where $\sigma
_{g,r}$ and $\sigma _{g,i}$ are real and imaginary parts of the graphene
conductivity, respectively, and  $\Delta $ is a very thin thickness. The
Kuba formalism in Ref. \cite{Hanson1, Hanson2} is applied for the graphene conductivity computation.
The initial input parameters are the temperature $T=3^{\circ }$K$%
,$ the relaxation time $\tau =1$ps, the ribbon width $w=200$nm, the graphene
thickness $\Delta =1$nm, and the frequency spectrum $28-32$THz. As shown,
the two first modes are the edge graphene surface plasmon modes (\textit{EGSPP})
propagating on the ribbon's edges. Third modes onwards, are the waveguide
graphene surface modes (\textit{WGSPP}) propagating on the surface of the
graphene ribbon. They arise from the surface plasmons. The number of modes and their propagation properties are tunable by the
ribbon width as well as the graphene parameters and frequency \cite{Nikitin}. In the following, the first waveguide mode (\textit{WGSPP1}) is only considered so that it is properly directed on the surface of the plasmonic bend structures. Taking into account that in the SPP modes dispersion diagram of a graphene ribbon the edge modes are separated from the waveguide modes with a wavenumber gap \cite{Nikitin}, we can ignore the edge modes. This feature is desirable in our design. The effective
refractive index $n_{\text{eff}}$ of the \textit{WGSPP1} mode at various
chemical potential $\mu_{c}$ for operating frequency $f=30$THz is extracted
and demonstrated in Fig. \ref{wgsp}b. Note that $n_{\text{eff}}=\beta /k_{0}$
where $\beta$ is the wavenumber of \textit{WGSPP1} and $k_{0}$ is
the free space wavenumber. The values are normalized by the effective index
for $\mu _{c}=150$meV; i.e. $n_{\text{eff}}(\mu_{c}=150$meV$)$ is
chosen as a reference value as explained later. As shown by increasing the
chemical potential, the effective refractive index is reduced.

\subsection{TO graphene-based bend and divider}

\bigskip In this section, two plasmonic bend structures are designed based
on graphene. A bend and a T-shaped divider are composed of graphene
having piecewise-constant regions of varying chemical potential, with the geometrical patterns shown in Fig. \ref{nTO}a,b. In the following, the graphene parameters of each region are determined to provide the required refractive index values obtained by TO. In other words, the
normalized $n_{\text{eff}}$ of WGSPP1 is considered to be equivalent to $n_{\text{TO}}$, ideally leading to the manipulation of the propagation direction of the
WGSPP1 mode on the surface of the proposal designs. It should be mentioned that multiplying the TO refractive index ($n_{\text{TO%
}}$) by a factor doesn't influence the ideal performance. Here, $n_{\text{eff}}$ of WGSPP1 is normalized instead. The normalization factor is the effective index of a region that contains the input port, whereas $\mu _{c}=150$meV. Furthermore, the region with $n_{\text{eff}}$ higher than one have negligible effect on performance comparing to the region with $n_{eff}<1$. It allows to use just a few discrete regions with $n_{eff}>1$ in the design.

\begin{figure}[h!]
\centering\includegraphics[width=10cm]{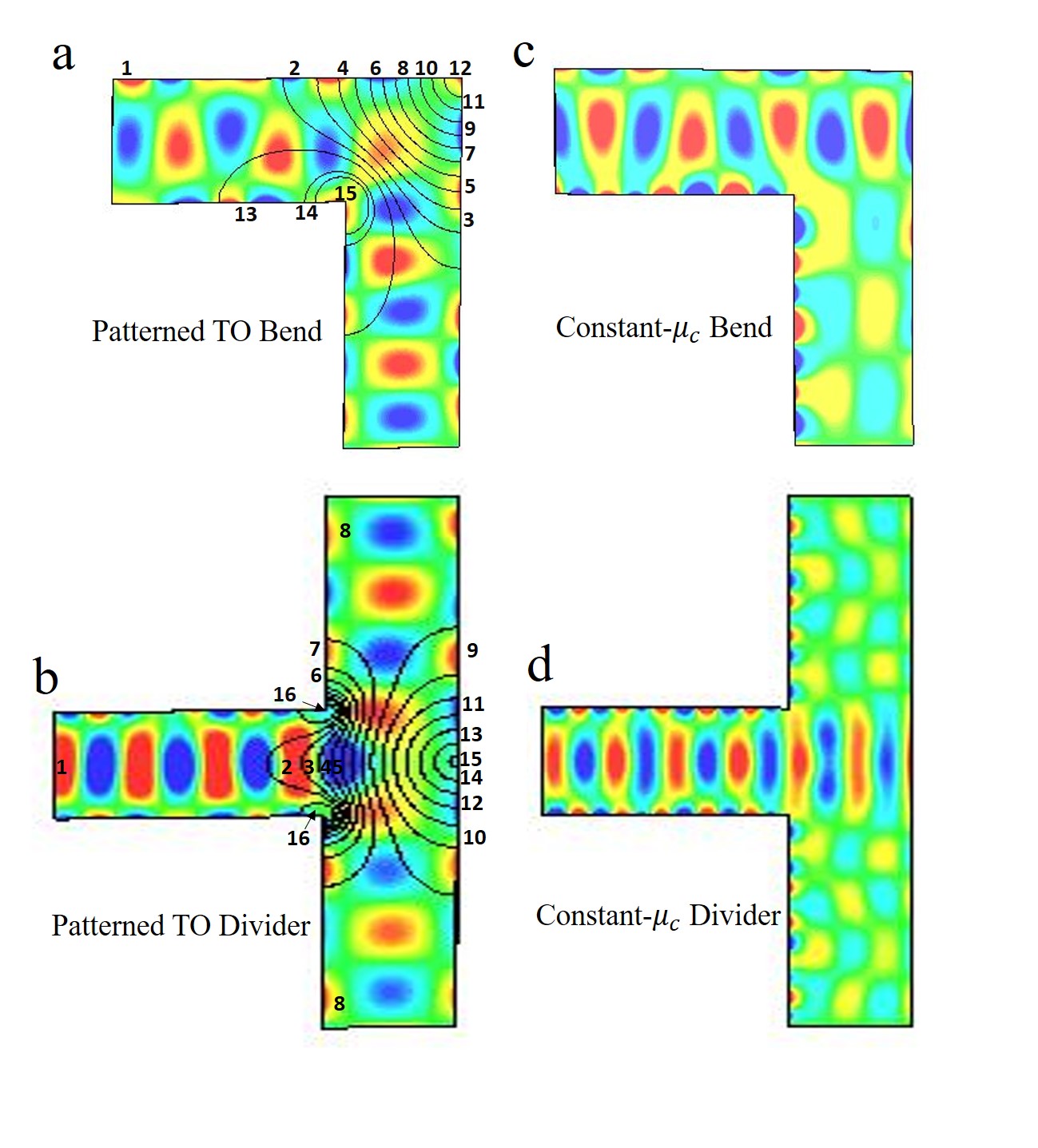}
  \caption{The electric field profile of the first waveguide surface mode on (a) a patterned graphene TO bend and (b) a patterned TO divider, and also on (c) a constant-$\mu_c$ bend and (d) a constant-$\mu_c$ T-shaped divider (d), in both cases with $\mu_c=150$meV.   
  }
 \label{wave}
\end{figure}

Considering $T=3^\circ$K, $\tau=1$ps for all regions, the only unknown parameter is the chemical potential. The $\mu_c$ values are extracted from the curve in Fig. \ref{wgsp}b to realize the required effective index in each region. The values are reported in Table \ref{T1} and \ref{T2} for a bend and a T-shaped divider, where $i$ corresponds to the number of regions shown in Fig. \ref{wave}a,b. The operating frequency is $f=30$THz. 
 
\begin{table}
\centering
\caption{The chemical potential of the graphene regions of a TO bend. i corresponds to the region numbers shown in Fig. \ref{wave}a. }
\label{T1}
 
\begin{tabular}{cccccc}
$i$ & $n_{\text{TO}},n_{\text{WGSP1,n}}$ & $\mu _{c} (\text{meV})$ & $i$ & $n_{\text{TO}%
},n_{\text{WGSP1,n}}$ & $\mu _{c} (\text{meV})$ \\ 
\hline
1 & 1 & 150 & 9 & 0.32 & 215 \\ 
2 & 0.9 & 158 & 10 & 0.24 & 220 \\ 
3 & 0.8 & 167 & 12 & 0.16 & 225 \\ 
4 & 0.73 & 174 & 13 & 0.08 & 228 \\ 
5 & 0.64 & 182 & 14 & 1.13 & 141 \\ 
6 & 0.56 & 191 & 15 & 1.21 & 136 \\ 
7 & 0.48 & 200 & 16 & 1.29 & 132 \\ 
8 & 0.40 & 207 & - & - & -%

\end{tabular}
\end{table}

\begin{table}
\centering
\caption{The chemical potential of the graphene regions of a TO T-junction. i corresponds to the region numbers shown in Fig. \ref{wave}b.}
\label{T2}
\begin{tabular}{cccccc}

$i$ & $n_{\text{TO}},n_{\text{WGSP1,n}}$ & $\mu _{c} (\text{meV})$ & $i$ & $n_{\text{TO}%
},n_{\text{WGSP1,n}}$ & $\mu _{c} (\text{meV})$ \\ 
\hline
1 & 1.00 & 150 & 9 & 0.43 & 205 \\ 
2 & 0.94 & 154 & 10 & 0.35 & 213 \\ 
3 & 0.86 & 160 & 12 & 0.28 & 218 \\ 
4 & 0.80 & 167 & 13 & 0.21 & 224 \\ 
5 & 0.72 & 175 & 14 & 0.14 & 227 \\ 
6 & 0.65 & 180 & 15 & 0.07 & 229 \\ 
7 & 0.58 & 190 & 16 & 0.035 & 230 \\ 
8 & 0.50 & 198 & 17 & 1.25 & 134%
 
\end{tabular}
\end{table}
Figure \ref{wave} shows the full wave simulation results. As shown in Fig. \ref{wave}a the surface plasmon polaritons are effectively guided from
the input port to the output port of a TO graphene bend. The surface wave doesn't reflect upon encountering the right-angled curve. 
Similarly, the electric field profile of the SPP in a patterned divider is
shown in Fig. \ref{wave}b. As shown, the surface waves pass through the right-angled junction without reflection and the wave is equally divided between two output ports. It behaves like an ideal plasmonic divider. 

In contrast to Fig. \ref{wave}a,b, the electric field profiles of a constant-$\mu_c$ bend and a constant-$\mu_c$ divider built from a single graphene region with $\mu _{c}=150$meV, are shown in Fig. \ref{wave}c,d. In comparison to the TO-designed bends, the ordinary bends show considerable reflections, as expected.
To have a quantitative comparison, we define a factor $\kappa$ as the ratio of power transmission coefficient of the patterned/unpatterned graphene bend and straight graphene waveguide. Note that the straight waveguide has the same length as the patterned/unpatterned graphene bend. For unpatterned bend (constant-$\mu_c$) the relative transmission factor is $\kappa=-15$dB, while for the patterned bend this value increases to $\kappa=-3$dB. Ideally $\kappa=0$dB, meaning that the wave propagation in the bend is the same as a straight waveguide, i.e. maximum power is transmitted from input to output port of the bend. For the TO-bend design, this value is close to the ideal case, which validates the accuracy of our model. Obviously, this value increases by increasing the number of graphene regions. For the divider we define the same relative transmission factor, assuming that the transmission coefficient of the ideal divider is one-half the transmission coefficient of the straight waveguide. For the unpatterned divider, the relative transmission factor is $\kappa=-23$dB. It becomes $\kappa=-5$dB for the patterned divider, which is a very significant improvement.

\subsection{Implementation of piecewise-constant graphene profile}

For a given chemical potential, the corresponding carrier density is determined using $
n_{e}=2/(\pi\hbar^{2}\upsilon_{f}^{2})%
%TCIMACRO{\dint \limits_{0}^{\infty}}%
%BeginExpansion
{\displaystyle\int\limits_{0}^{\infty}}
%EndExpansion
\epsilon(f_{d}(\epsilon)-f_{d}(\epsilon+2\mu_{c}))d\epsilon$
where $\hbar $ is the reduced Plank constant, $e$ is the electron
charge, $\upsilon _{f}$ is the Fermi velocity, and $f_d$ is the Fermi energy defined as  $f_{d}(\epsilon)=1/(e^{(\epsilon-\mu_{c})/k_{B}T}+1)$. For chemical potential values reported in Table \ref{T1} and \ref{T2},  the range of variation of carrier density $n_e$ is $[1.32-3.3] \times 10^{12} (\text{cm}^{-2})$.

\begin{figure}[h!]
\centering\includegraphics[width=12cm]{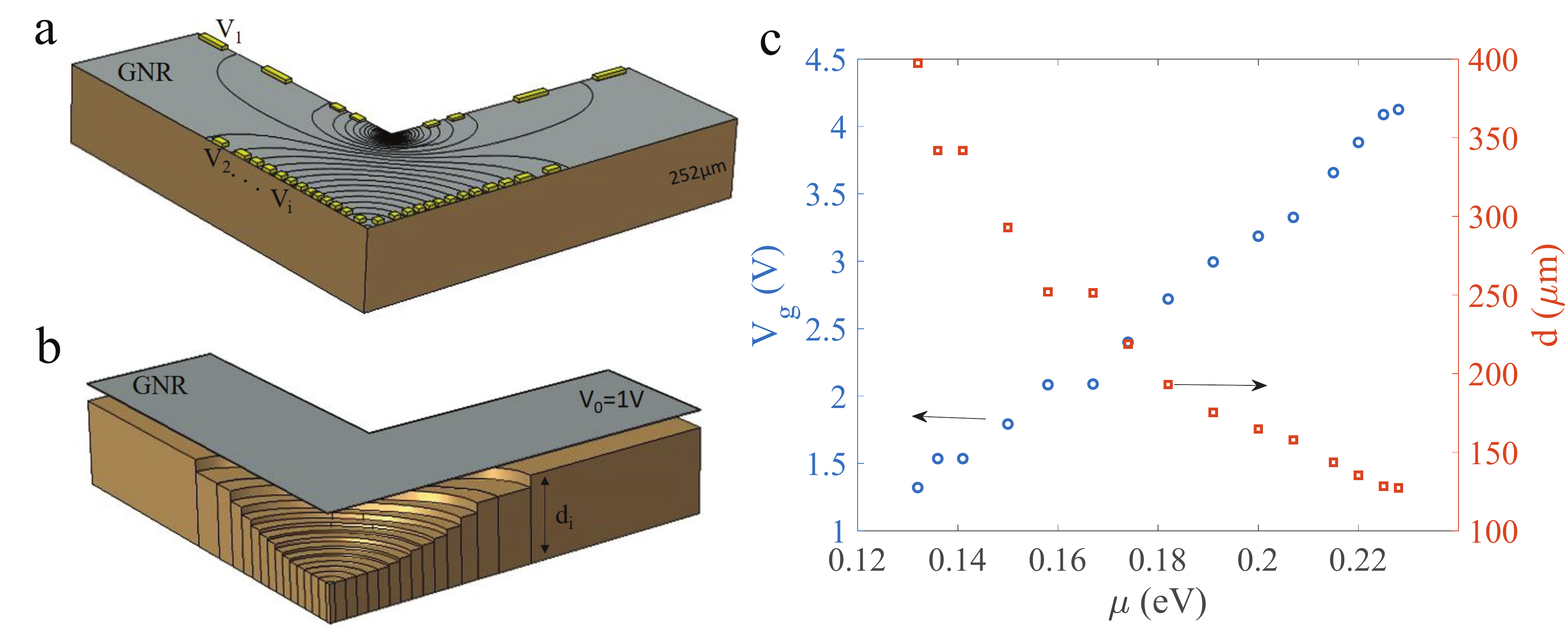}
  \caption{ Parameters of two fabrication schemes a. Applying multiple gates with uniform substrate of 525$\mu$m. The voltage of a gate connected to each graphene regions is shown in Panel c (blue curve). b. Applying uneven substrate underneath a uniform graphene bend, wherein a constant voltage $V_0$=1V is applied. The substrate height is shown in Panel c (red curve) }     
  
 \label{vd}
\end{figure}

The carrier density in a graphene nano-ribbon can be controlled by voltage or chemical doping, or by applying an uneven ground plane underneath the graphene layer, which results in a nonuniform carrier density and chemical potential on the surface of graphene \cite{Vakil1}. The variation of the separation distance between graphene and the metal plane can implement the chemical potential profile \cite{Ebr1, Ebr2}. In a graphene ribbon the ratio of the bias voltage ($V_g$) and the substrate height ($d$) satisfies the relation $V_g/d=e(e\mu _{c})^{2}/\epsilon\pi\hbar ^{2}\upsilon_{F}^{2}$, where $\epsilon$ is the substrate permittivity \cite{Ebr1}. To implement the nonuniform chemical potential profile obtained for the bend and divider in Fig. \ref{wave}a,b, we propose two fabrication methods. One is to use a constant-thickness substrate and print a disjointed ground plane, with each ground-plane segment having the geometry of the corresponding graphene region shown in Fig. \ref{wave}. Individual voltages are applied to each ground plane segment to implement the correct chemical potential (Fig. \ref{vd}a). Consider a quartz/Silicon substrate with total thickness $d=525\mu$m, the required voltage for each graphene region is calculated and shown in Fig. \ref{vd}c (blue). As shown, it varies from $[1-4.5]$V. However, in this case the graphene regions need to be separated with a small gap.
%, which increases the stored energy and reduces the efficiency.}  
In the second proposed implementation, an uneven (variable-height) ground plane is utilized underneath a continuous graphene bend (see Fig. \ref{vd}b). By considering a fixed voltage $V_g=1$V, the substrate thickness under each graphene region is determined using the above relation and shown in Fig. \ref{vd}c (red). The resulting thickness of substrate varies from 100$\mu$m to 400$\mu$m.

\bigskip 

\section{CONCLUSION}

In summary, the transformation media of a THz bend and a THz T-shaped divider were realized using graphene. The ribbon graphene SPP is properly guided on the surface of the two proposed bend structures. The surface wave can travel through 90 degree bends with no refection. The relative transmission coefficient of TO-designs is significantly larger than the components with constant chemical potential. The required voltage and substrate thickness were calculated for implementation of the obtained piecewise-constant graphene profiles.  

\section*{Acknowledgement}

Funding for this research was provided by the National Science Foundation
under grant number EFMA-1741673.

%\section*{Data Availability}

%The datasets generated during and/or analyzed during the current study are available from the corresponding author on reasonable request.

\section*{Disclosures}
 The authors declare no conflicts of interest.

\end{document}